\documentclass[pdflatex,sn-mathphys-ay]{sn-jnl}

\usepackage{graphicx}%
\usepackage{multirow}%
\usepackage{amsmath,amssymb,amsfonts}%
\usepackage{amsthm}%
\usepackage{mathrsfs}%
\usepackage[title]{appendix}%
\usepackage{xcolor}%
\usepackage{textcomp}%
\usepackage{manyfoot}%
\usepackage{booktabs}%
\usepackage{algorithm}%
\usepackage{algorithmicx}%
\usepackage{algpseudocode}%
\usepackage{listings}%
\usepackage[T1]{fontenc}
\usepackage[latin9]{inputenc}
\usepackage{hyperref}

\theoremstyle{thmstyleone}%
\theoremstyle{thmstyletwo}%

\theoremstyle{thmstylethree}%
\begin{document}

\title[Mode Collapse in Nested Sampling]{Mode Collapse in Nested Sampling}

\author*[1]{\fnm{Johannes} \sur{Buchner}}\email{johannes.buchner.acad@gmx.com}

\affil*[1]{\orgname{Max Planck Institute for Extraterrestrial Physics},
\orgaddress{\street{Giessenbachstrasse}, \city{Garching},
\postcode{85748}, \country{Germany}}}

\abstract{
Nested Sampling is a Monte Carlo algorithm enabling posterior estimation
and Bayesian model comparison, and is especially robust in multi-modal posteriors.
This is because nested sampling maintains a population
of live points sampled from the entire prior.
In each iteration, the population is advanced above a likelihood threshold, potentially discarding
modes ruled out by the data. However, the Monte Carlo nature of point replenishment
can also accidentally discard a mode.
We draw a connection to the neutral Moran process in genetics,
and quantify the occurrence probability of this failure mode of nested sampling
with a simple symmetric random walk model on the live point occupancy.
We find a simple rule for setting the minimum number of live points
so that mode die-out is made unlikely.
}


\maketitle

\section{Introduction}

In the physical sciences, Skilling's nested sampling \citep{Skilling2004,skilling2006nested}
is a popular Monte Carlo algorithm to achieve Bayesian model comparison
and posterior inference. It is especially robust in multi-modal posteriors \citep{Buchner2023}, and this paper shows additional evidence of this.

Nested sampling maintains a population of $K$ live points sampled from the prior.
In each iteration, the point with the lowest likelihood is discarded, and replaced
by a new prior sample above that likelihood threshold (likelihood-restricted prior sampling, LRPS).
Thus, in each iteration, prior volume where the likelihood is below this threshold becomes inaccessable.
The fraction of the prior probability mass excluded can be statistically estimated from order statistics.
These replaced points across iterations allow estimation of the evidence and posterior probability distribution \citep[see, e.g.][for a pedagogical introduction]{Ashton2022}.
The estimation is based on a one-dimensional integral approximated by a weight sum of the volume restrictions at each iteration times the likelihood of the replaced point.
This work focuses on the faithfulness of likelihood-restricted prior sampling in multi-modal settings.

In high-dimensional settings, the replacement is achieved
by running a Markov chain started from a randomly chosen live point
\citep[see][for a review of variations]{Buchner2023}.
We focus here on such algorithms that generate new samples
by starting a proposal from a randomly selected live point. This class includes
step samplers based on hit-and-run or slice sampling, and some region-based algorithms
such as MLFriends \citep{Buchner2016,Buchner2019c}.
In multi-modal states, the discarded point is removed from one mode, and
a replacement point is generated in a mode sampled proportional to the live point occupancy.
It is unlikely that the Markov chain succeeds to traverse the disconnected modes.

As one example for multi-modality, the velocity time series analyses for the discovery of exoplanets
with nested sampling \citep[e.g.,][]{Nelson2020}
can exhibit modes corresponding to the planet period and higher-order aliased periods.
Intermediate periods are ruled out at intermediate likelihood thresholds, i.e., early in the nested sampling run.
The orbital parameter ranges allowed for each period (alias) may differ,
corresponding to vastly different mode weights at any nested sampling iteration ($p\ll50\%$).

\section{Method}
We analyse the probability of unintentional mode die-out.
We being by making a number of assumptions, followed by introducing
a simple two-mode state model, that is probabilistically updated 
in each nested sampling iteration. We then study the behaviour of this
modelling of the nested sampling process.

\subsection{Assumptions}
\label{sec:assumptions}

Firstly, we assume that there are two modes (0 and 1), with fractional constrained prior mass
$p\in(0,1)$ and $1-p$, respectively, at the likelihood level under consideration.
The modes are populated by $K$ live points in total.
The analysis presented here is for two modes; extension to more modes is left for future work.
Secondly, we assume the probability ratio $p/(1-p)$ remains approximately fixed
over the nested sampling iterations of interest; this is a local approximation
valid at any fixed likelihood level and is relaxed in the Discussion.
Thirdly, we assume the likelihood-restricted prior sampler (LRPS) is faithful, i.e.,
new points are drawn proportional to the remaining constrained prior mass in each mode.
Fourthly, we assume mode rediscovery is negligible,
i.e., once all live points leave a mode, they do not return.

For our analysis, we define mode collapse as a loss of either mode
merely due to the random replacement sampling.
Our analysis is independent of sampling dimension.

\subsection{State model and initialisation}
\label{sec:init}
Our model tracks the number of live points $i$ that are members of the first mode 
(and $K-i$ are in the second mode).
At a given likelihood level, there are $K+1$ possible states,
ranging from $i=0$ to $i=K$ live points in mode 1 (and the remainder in mode 0).
The initial state probabilities follow a binomial distribution:
\begin{equation}
\mathbb{P}(X_0 = i) = \binom{K}{i} p^i (1-p)^{K-i}, \quad i=0,1,\dots,K.
\label{eq:initial}
\end{equation}
This models the random occupancy of the two modes at the likelihood level first considered.
It is exact if the live points at that level are independent draws from the constrained prior,
which holds at the very first nested sampling iteration (direct prior sampling) and remains
a good approximation at later iterations provided the sampler has been faithful up to that point.
Note that for very small $p$, the dominant risk is non-discovery of the small mode at
initialisation, i.e.\ $\mathbb{P}(X_0=0)\approx(1-p)^K$, rather than subsequent stochastic die-out;
the evolved collapse probability $\mathbb{P}_{\mathrm{evol}}(t)$ defined below conditions on
both modes being present at initialisation and therefore addresses the two risks separately.

\subsection{Selecting a live point for removal}
In each nested sampling iteration, the lowest-likelihood live point is removed and a
replacement point is generated above the threshold.
We thus need to consider the probability that the removed point
(the lowest-likelihood live point) belongs to one mode or the other.
Naively, one may expect this to be proportional to the prior volume of the mode it belongs so, 
but the probability of being selected for removal depends on the evolution of the likelihood threshold.

The mode with a shallower likelihood gradient near the current threshold loses points
preferentially, because its live points tend to be spread across lower likelihood values.
Such a mode is at genuine risk of die-out, but this is not a failure of nested sampling.
It may simply be a less favoured mode being correctly eliminated by the data.

The scientifically important failure mode is the accidental loss of a mode that
\emph{should} be retained: one whose live points are concentrated above the threshold,
i.e., whose likelihood rises steeply. For such a mode, the probability of being
selected as the lowest-likelihood point is \emph{lower} than its fractional occupancy $i/K$
would suggest. The replacement step, however, is unaffected by likelihood gradients:
the seed is drawn uniformly from all $K$ live points, so the probability that the
replacement point is assigned to mode 1 remains $i/K$ regardless of gradient.
Together, the net probability that mode 1 (the steep-gradient mode) loses a point
per iteration is \emph{less} than under the symmetric model, because it is less likely
to supply the removed point while equally likely to supply the replacement seed.
Using $i/K$ as the removal probability therefore
\emph{overestimates} how often this mode loses a point on net, giving a conservative
(upper-bound) estimate of its die-out risk.

We therefore proceed under the assumption of equal likelihood gradients in both modes,
which gives a conservative estimate of the number of live points needed to make
accidental die-out of a mode that should be retained unlikely.

\subsection{Update}
In the LRPS algorithms we consider, the replacement point's mode is inherited 
from the seed live point (the seed of the Markov chain).
In the implementation class considered here, the dead point remains eligible
to serve as the seed, since it still lies on the current likelihood contour 
and can validly seed a proposal into the constrained region. The seed is therefore
drawn uniformly from the full pre-update population of K live points.
Under the equal-gradient assumption of Section~\ref{sec:assumptions}, and conservatively
for the steeper-gradient mode, the probability that the removed point belongs to mode 1
is $i/K$. The probability that the replacement point is assigned to mode 1 ---
determined by the mode of the randomly chosen seed --- is also $i/K$,
since the seed is drawn uniformly from the current $K$ live points.
These two draws are independent (both sampled from the pre-update population
of $K$ points), so the number of live points in mode 1 changes by
$+1$, $-1$, or $0$ with probabilities:
\begin{align}
\mathbb{P}(X_{t+1} = i+1 \mid X_t = i) &= \left(1-\frac{i}{K}\right)\frac{i}{K}, \label{eq:trans_up}\\
\mathbb{P}(X_{t+1} = i-1 \mid X_t = i) &= \frac{i}{K}\left(1-\frac{i}{K}\right), \label{eq:trans_down}\\
\mathbb{P}(X_{t+1} = i \mid X_t = i) &= \left(\frac{i}{K}\right)^2 + \left(1-\frac{i}{K}\right)^2,
\label{eq:trans_stay}
\end{align}
for $i=0,1,\dots,K$.

Note that the up and down transition probabilities are equal, so this is a
\emph{symmetric} random walk on $\{0,1,\ldots,K\}$, analogous to the neutral Moran process
(Section~\ref{sec:moran}). The chain is lazy (the probability of remaining at the same state
is always at least $1/2$, achieving this minimum at $i=K/2$ and equalling $1$ at the
absorbing boundaries $i=0$ and $i=K$) because the removal and replacement draws are independent.
States $0$ and $K$ are absorbing: once all live points are in one mode,
the faithful sampler can only produce points in that mode.
All other states are transient, and the chain is absorbed almost surely as $t\to\infty$.
We are interested in the finite-time probability of absorption over the iterations
relevant for evidence accumulation.

Note that absorption at $i=K$ (all live points in mode 1, mode 0 eliminated) is also
counted as premature mode collapse in our definition, since the goal is to retain both modes
throughout the period of the run while the volume ratio stays constant. 
In the symmetric equal-gradient model, both absorbing states are
equally likely to be reached from symmetric starting states. For unequal $p$, the
smaller mode is at greater risk, and the conservative upper bound on its die-out
probability is our primary quantity of interest.

\subsection{Numerical evaluation}
\label{sec:method-numerical}
Starting from the initial probability vector given by Eq.~\ref{eq:initial},
we multiply by the transition matrix defined by
Eqs.~\ref{eq:trans_up}--\ref{eq:trans_stay}
repeatedly ($t$ times) to obtain the probability distribution 
of each possible state at iteration $t$. 
This power of matrices of size $(K+1,K+1)$ is cheap to compute for moderate $K$.

\subsection{Collapse probability}
We define the probability of mode collapse at iteration $t$ as:
\begin{equation}
\mathbb{P}_c(t) = \mathbb{P}(X_t = 0) + \mathbb{P}(X_t = K).
\label{eq:probcollapse}
\end{equation}
We separate out the probability of collapse at initialisation:
\begin{equation}
\mathbb{P}_c(0) = (1-p)^{K} + p^K,
\label{eq:initialcollapse}
\end{equation}
which corresponds to one mode being unoccupied from the very first iteration
due to random sampling. The conditional probability of collapse by iteration $t$,
given that both modes were initially populated, is then:
\begin{equation}
\mathbb{P}_{\mathrm{evol}}(t) = \frac{\mathbb{P}_c(t) - \mathbb{P}_c(0)}{1 - \mathbb{P}_c(0)}.
\label{eq:probcollapsefinal}
\end{equation}

\subsection{Connection to the neutral Moran process}
\label{sec:moran}
Our model is analogous to the neutral Moran process in population genetics \citep{moran1958random}.
A population of $K$ individuals carries a gene that exists in two allelic states.
In each generation, one individual is chosen uniformly at random to die,
and one individual (possibly the same) is chosen uniformly at random to reproduce,
keeping the population size fixed at $K$.
In the neutral case, allelic state does not affect fitness.
A newly introduced allele will either go extinct or reach fixation (spread to all $K$ individuals),
with the fixation probability from state $i$ equal to $i/K$ \citep{moran1958random}.

The mean time to absorption (fixation or extinction) from state $i$ is
\citep{moran1958random,Watterson1961}:
\begin{equation}
k_{i}=K\left[\sum_{j=1}^{i}\frac{K-i}{K-j}+\sum_{j=i+1}^{K-1}\frac{i}{j}\right],
\label{eq:meanabsorptiontime}
\end{equation}
with $k_0 = k_K = 0$ for the two absorbing states.
Here the first sum accounts for paths to extinction (state 0)
and the second for paths to fixation (state $K$),
weighted by the current occupancy $i$.
Marginalising over the initial distribution (Eq.~\ref{eq:initial}) gives
the mean absorption time averaged over all starting configurations:
\begin{equation}
k(p,K) = \sum_{i=0}^{K} \mathbb{P}(X_0 = i)\, k_{i}.
\label{eq:meanabsorptiontimefinal}
\end{equation}

\section{Results}

\begin{figure}
\includegraphics[width=\columnwidth]{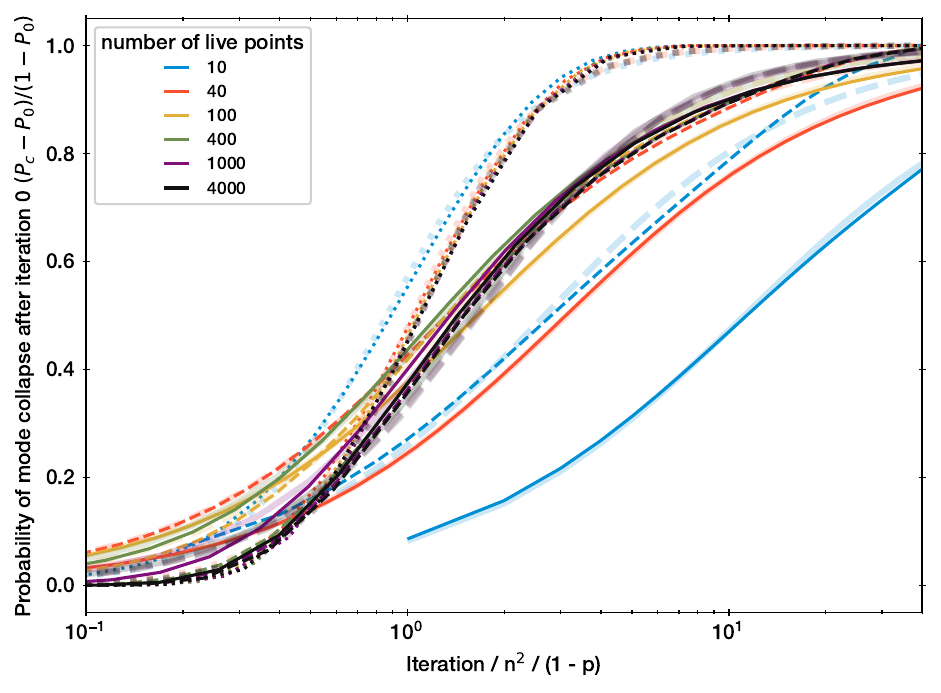}
\caption{Probability of evolved mode collapse $\mathbb{P}_{\mathrm{evol}}(t)$
(Eq.~\ref{eq:probcollapsefinal}) as a function of iteration number.
The x-axis is normalised by $K^2 \times \min(p,1-p)$,
for $p=0.5$ (dotted curves), $p=0.05$ (dashed curves), and $p=0.01$ (solid curves).
The y-axis is the conditional collapse probability given that both modes were
populated at initialisation.
All curves have a sigmoid functional form.
The 50\% collapse probability is reached between 1 and 3 on the normalised x-axis
across all tested values of $p$ and $K$, a factor-of-three spread in location.
}
\label{fig:modecollapseevol}
\end{figure}

Figure~\ref{fig:modecollapseevol} shows the probability of mode collapse
as a function of the number of nested sampling iterations, for several scenarios.
This result is obtained through numerical evaluation (Section~\ref{sec:method-numerical}).
We find that plotting against $\log$ iteration yields an approximately 
sigmoidal rise as a function of $\log$ iteration,
as presented in Figure~\ref{fig:modecollapseevol}.
These results are obtained from sequences of iterations for
three values of $p$ (0.5, 0.05, 0.01) and four values of $K$ (40, 100, 400, 1000).
Across $p$ and $K$, curves are well-aligned when the x-axis is normalised by
$K^2 \times \min(p,1-p)$; we note that this scaling is empirically observed from
the numerical results rather than analytically derived.
There is some variation in slope, but a general trend is clear:
the probability of evolved mode collapse remains below 50\% when
$\mathrm{iteration} \lesssim K^2 \times \min(p,1-p)$.
Normalising by the mean absorption time $k(p,K)$ (Eq.~\ref{eq:meanabsorptiontimefinal})
instead gives more diverse intercepts and is a less useful scaling.

We can identify the total number of nested sampling iterations from prior to posterior
as approximately $K \times H$ \citep{skilling2006nested},
where $H$ is the information gained in updating from prior to posterior (in nats),
equal to the Kullback--Leibler divergence of the posterior from the prior.
Setting the iteration budget $KH \lesssim K^2 \times \min(p,1-p)$ and solving gives
a heuristic rule of thumb for the minimum number of live points:
\begin{equation}
K \gtrsim \frac{H}{\min(p,1-p)}.
\label{eq:recommendation}
\end{equation}
Since the 50\% collapse threshold used to derive the scaling is not a stringent
safety criterion, in practice one should aim for $K$ to comfortably exceed this lower bound,
i.e., $K \gg H/\min(p,1-p)$, with the required margin readable from the sigmoid in
Figure~\ref{fig:modecollapseevol}: achieving a collapse probability well below 50\%
requires $K$ to exceed the bound by a factor of a few.
That is, the number of live points should exceed the KL divergence of the posterior
from the prior divided by the fractional constrained prior mass of the smallest mode
at the relevant likelihood levels.

\section{Discussion}
In this paper, we considered how the size of the live point population impacts retention of modes
in nested sampling. It is intuitive that with few live points, random replacement creates the
risk of mode die-out after a number of iterations.
Our result is quite general. While based on numerically evaluated matrix multiplications, 
our results are based on an analytic framework that is independent of dimensionality and 
the point replacement algorithm, assuming lost modes cannot be rediscovered.

Our main result (Eq.~\ref{eq:recommendation}) provides a heuristic scaling law for the
minimum number of live points that makes mode die-out unlikely:
the number of live points should be set to at least
the KL divergence of the posterior from the prior (in nats) divided by the fractional constrained
prior mass of the smallest mode at the relevant likelihood levels.
With a typical information gain of $H \sim 10$--$1000\,\mathrm{nats}$ as discussed
in \cite{skilling2006nested},
and a fiducial smallest-mode fractional constrained prior mass of $p=0.1$,
Eq.~\ref{eq:recommendation} gives a recommendation of order $100$--$10000$ live points.
Assuming that multi-modality is relevant for only a fraction of the $KH$ iterations
(e.g., 5\%), the effective requirement is reduced by that factor,
bringing the estimate to order $10$--$500$ live points.
This is consistent with typically used values in the literature,
and suggests that accidental mode die-out is exceedingly rare under commonly used configurations.
After an initial nested sampling run, eq.~\ref{eq:recommendation} can estimate the probability
of unintentional mode die-out.

The number of live points is one of the most fundamental parameters in nested sampling,
setting the sampling resolution and thus the probability of discovering a mode (section~\ref{sec:init}),
and the number of nested sampling iterations for a given information gain~\citep{Skilling2004,evans2007discussion,Chopin2010}.
Aside from the theoretical cost of evidence accumulation,
the practical proposal construction and its efficiency are also relevant.
Counterintuitively, \cite{Buchner2023} discussed that in many practical nested sampling algorithms,
increasing the number of live points increases sampling efficiency, because the proposal
built from the $K$ live points becomes more refined with increased $K$,
improving acceptance probability.

Dynamic mode-die-out prevention methods have been proposed.
\cite{Feroz2008} developed an explicit approach to prevent mode die-out.
When their clustering of live points detects well-separated modes, they
determine the relative importance of the prior probability mass of each mode by Monte Carlo integration,
and start a nested sampling run for each cluster sub-space.
The sub-runs and the main run are then combined to give the total Bayesian evidence,
the posterior masses for each mode, and posterior samples.
This introduces additional implementation complexity but avoids the variable number of live points,
especially if the modes are of very different fractional constrained prior mass $p\ll1$.
This behaviour is implemented in MultiNest (but disabled when importance nested sampling is enabled).
As part of their reactive nested sampling extension of dynamic nested sampling,
\cite{Buchner2021c} proposed increasing the number of live points
when multiple modes are detected, until each mode is populated with at least
a number $K_\mathrm{min}$ of live points. This behaviour is implemented in UltraNest~\citep{Buchner2021ultranest}.
This strategy would also likely make mode die-out as analysed here unlikely.
Both of the above approaches hinge on clustering being able to detect modes.
One may suspect that if separated modes cannot be detected by the clustering algorithm,
a Markov chain may also not be hindered from traversing across modes;
however, this cannot be guaranteed in general.

\backmatter


\bmhead{Acknowledgements}

We thank Andrew Fowlie for feedback on the manuscript.
This research has built upon large language models (LLM):
The mode collapse research question was stated to ChatGPT 5 in an abstract way, 
which revealed the connection to Markov transition analysis.
Initial scaffolding of the numerical simulation code was also made with the same LLM.
The final code was written by the author.
The paper was written by hand. However, ChatGPT 5.4 and Anthropic Claude 4.6 
were used to identify issues with math and grammar.
The connection to neutral Moran processes was also identified by the LLM. 
The scienceOS LLM platform to search for references on Moran processes.

%
%

\bibliography{agn,stats}
\end{document}